\documentclass[journal=jcim,manuscript=article]{achemso}

\newcommand{\titleName}[0]{MIST-CF: Chemical formula inference from tandem mass spectra}
\newcommand{\methodName}[0]{MIST-CF}
\newcommand{\methodNameFull}[0]{MIST-CF: Metabolite Inference with Spectrum Transformers for Chemical Formulae}
\newcommand{\gnpsData}[0]{NPLIB1}
\newcommand{\nistData}[0]{NIST20}

\newcommand{\codeUrl}[0]{\url{https://github.com/samgoldman97/mist-cf}}
\newcommand{\codeUrlNist}[0]{\url{https://github.com/samgoldman97/nist-parser}}

\newcommand{\zenodoID}[0]{8151490}
\newcommand{\zenodoIDGithub}[0]{8151513}

\newcommand{\mol}{\mathcal{M}}
\newcommand{\spec}{\mathcal{S}}
\newcommand{\numAnnotated}{N_p}

\newcommand{\inten}[1]{I}
\newcommand{\intenIdx}[1]{I^{(#1)}}
\newcommand{\mass}{M}
\newcommand{\massIdx}[1]{M^{(#1)}}

\newcommand{\prodFormIdx}[1]{\bm{f}^{(#1)}}

\newcommand{\preForm}{\bm{\mathcal{F}}}
\newcommand{\preFormIdx}[1]{\preForm_{(#1)}}

\newcommand{\preAdduct}[0]{\bm{\mathcal{A}}}
\newcommand{\preAdductIdx}[1]{\preAdduct_{(#1)}}
\newcommand{\prodAdduct}[0]{\bm{a}}
\newcommand{\prodAdductIdx}[1]{\prodAdduct^{(#1)}}
\newcommand{\prodFormError}[1]{\epsilon^{(#1)}}

\newcommand{\modelGeneral}[0]{G_{\theta}}
\newcommand{\modelMIST}[0]{G^{(\text{MIST})}_{\theta}}
\newcommand{\fastFilter}[0]{FastFilter}
\newcommand{\modelFast}[0]{G^{(\text{FF})}_{\theta}}
\newcommand{\modelFFN}[0]{G^{(\text{FFN})}_{\theta}}
\newcommand{\modelXformer}[0]{G^{(\text{TF})}_{\theta}}

\newcommand{\enc}[1]{%
    \ifthenelse{\equal{#1}{}}%
    {\text{Enc}}%
    {\text{Enc}(#1)}
}

\newcommand{\hiddenState}[1]{%
    \ifthenelse{\equal{#1}{}}%
    {\bm{h}}%
    {\bm{h}^{(#1)}}
}

\newcommand{\hiddenStatePrime}[1]{%
    \ifthenelse{\equal{#1}{}}%
    {\bm{\tilde{h}}}%
    {\bm{\tilde{h}}^{(#1)}}
}

\newcommand{\MLP}[1]{%
    \ifthenelse{\isempty{#1}}%
    {\textsf{MLP}}%
    {\textsf{MLP}{(#1)}}
}

\newcommand{\transformer}{\textsf{Transformer}}

\newcommand{\realsPosCt}[1]{\mathbb{R}^{#1+}}

\newcommand{\intsPos}{\mathbb{Z}^{+}}

\newcommand{\indicator}[1]{
  \mathbb{I}({#1})
}

\newcommand{\specContext}{\bm{c}}

\usepackage{amsmath,amsfonts,bm}

\def\eqref#1{equation~\ref{#1}}

\def\1{\bm{1}}

\DeclareMathAlphabet{\mathsfit}{\encodingdefault}{\sfdefault}{m}{sl}
\SetMathAlphabet{\mathsfit}{bold}{\encodingdefault}{\sfdefault}{bx}{n}

\usepackage{soul}
\usepackage{chemformula} %
\usepackage[T1]{fontenc} %
\usepackage{caption}
\usepackage{subcaption}
\usepackage{graphicx}               %
\usepackage{amsmath,amssymb}        %
\usepackage{tabularx,booktabs}
\usepackage{pifont}
\usepackage{todonotes}
\usepackage{xifthen}
\usepackage{mathtools}
\usepackage{mhchem}

\usepackage{nameref}

\author{Samuel Goldman}
\affiliation{Computational and Systems Biology, Massachusetts Institute of Technology, 
Cambridge, Massachusetts 02139, USA}
\altaffiliation{Equal contribution}

\author{Jiayi Xin}
\affiliation{Statistics and Actuarial Science, The University of Hong Kong, Hong Kong 999077, China}
\altaffiliation{Equal contribution}

\author{Joules Provenzano}
\affiliation[ChemE]{Department of Chemical Engineering, Massachusetts Institute of Technology, 
Cambridge, Massachusetts 02139, USA}

\author{Connor W. Coley}
\affiliation[ChemE]{Department of Chemical Engineering, Massachusetts Institute of Technology, 
Cambridge, Massachusetts 02139, USA}
\alsoaffiliation{Department of Electrical Engineering and Computer Science, Massachusetts Institute of Technology, 
Cambridge, Massachusetts 02139, USA}
\email{* ccoley@mit.edu}

\title{\titleName{}}

\begin{document}




\begin{abstract}
Chemical formula annotation for tandem mass spectrometry (MS/MS) data is the first step toward structurally elucidating unknown metabolites. While great strides have been made toward solving this problem, the current state-of-the-art method depends on time-intensive, proprietary, and expert-parameterized fragmentation tree construction and scoring. In this work we extend our previous spectrum Transformer methodology into an energy based modeling framework, \methodNameFull{}, for learning to rank chemical formula and adduct assignments given an unannotated MS/MS spectrum. Importantly, \methodName{} learns in a data dependent fashion using a Formula Transformer neural network architecture and circumvents the need for fragmentation tree construction. We train and evaluate our model on a large open-access database, showing an absolute improvement of 10\% top 1 accuracy over other neural network architectures. We further validate our approach on the CASMI2022 challenge dataset, achieving nearly equivalent performance to the winning entry  within the positive mode category without any manual curation or post-processing of our results. These results demonstrate an exciting strategy to more powerfully leverage MS2 fragment peaks for predicting MS1 precursor chemical formula with data driven learning. 
\end{abstract}


\section{Introduction}
The discovery of previously unknown small molecules in biological samples is rapidly expanding our knowledge of plant chemistry \cite{pluskal_2019_biosynthetic, torrens-spence_pbs3_2019}, cancer biology \cite{cao_2022_commensal, dang_2009_cancer}, host-microbiome interactions \cite{quinn2020global, paik_2022_human, sato_2021_novel}, and other metabolite-mediated human biology \cite{wishart_metabolomics_2019}. Similar small molecule discoveries in the environmental sciences have led to new insights regarding the exposome and pollutant effects, resolving mysteries such as high salmon mortality rates.\cite{bundy_environmental_2009, tian_2021_ubiquitous} Increasing our ability to detect and identify the so-called ``dark metabolome'' with analytical chemistry techniques represents an exciting opportunity in experimental and computational chemistry.

Tandem mass spectrometry (MS/MS) is a particularly well-suited analytical technique for this, as it allows for the high throughput characterization of small molecules from complex mixtures. \cite{neumann_computational_2010} In an MS/MS experiment, both an intact ionized mass (MS1) and a set of fragment peak masses (MS2) can be measured for each unknown molecule.
This fragmentation spectrum serves as a structural representation of the molecule, ideally allowing the practitioner to match a small molecule structure to each resulting spectrum based upon database spectra matches. Due to the vastness of chemical space, however, many observed spectra have no precedent; in a large public MS/MS database, 87\% of observed spectra remain unannotated  \cite{bittremieux2022critical}. 

\begin{figure}[H]%
\centering
\includegraphics[width=\textwidth]{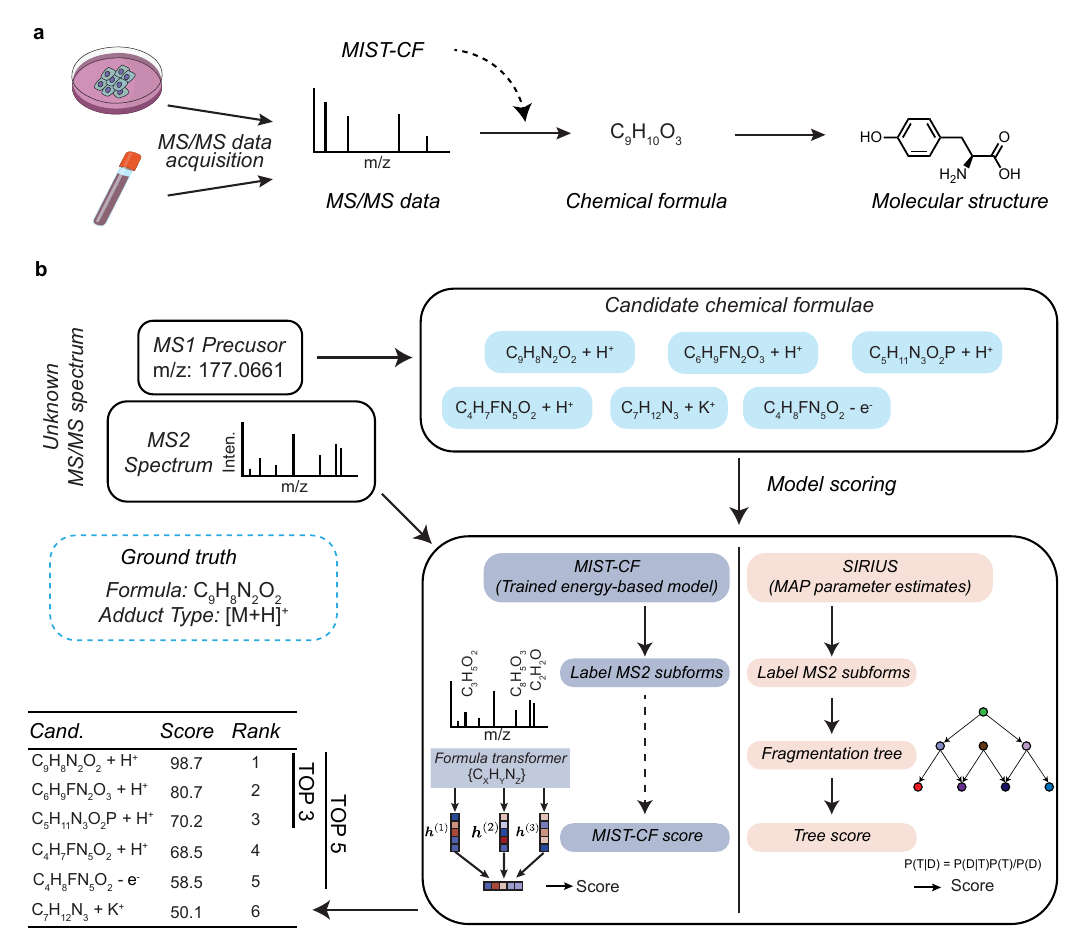}
\caption{\textbf{MIST-CF and SIRIUS both address the MS/MS chemical formula annotation problem.} 
\textbf{a.} Input samples are first processed, recording a tandem mass spectrum. Before assigning full molecular structures, the chemical formula can first be inferred to constrain structure assignment. \textbf{b.} Methodological similarities and differences between \methodName{} and SIRIUS.  A candidate precursor mass is first decomposed into plausible chemical formulae and adduct pairs. \methodName{} (left) learns in a data-driven fashion to assign scores and circumvents the need of fragmentation tree construction compared to SIRIUS (right).}\label{fig:overview}
\end{figure}

In such instances without spectral matches, we must rely on informatics and predictive modeling to identify the likely molecular structure. This pipeline almost always begins by inferring a chemical formula from the observed spectrum (Figure \ref{fig:overview}a). 
There are many formula options for each observed MS1 value, especially for higher mass compounds. 
Specifying the chemical formula (e.g.,  \ch{C6H12O6}, \ch{C9H11NO3
}, etc.) constrains the space of potential compound candidates to a set of isomers, whereas the MS1 alone only constrains the space of candidates to those with similar masses. %
Automated assignment is far from trivial; while the the maximum formula annotation accuracy was 94\% in in the recent Critical Assessment of Metabolite Identification in 2022 (CASMI20220),\cite{noauthor_critical_nodate} the median score was 71\%, \emph{and} these percentages were calculated only for the submitted predictions, rather than the total number of spectra tested. Improving the automated prediction accuracy of this step  promises to improve, simplify, and speed up downstream analyses. 

Chemical formula annotation tools can be grouped into two categories: database dependent  and database independent. Database dependent searches place restrictions on potential formulae, querying the candidate mass and spectrum against databases including NIST \cite{noauthor_tandem_nodate}, GNPS \cite{wang_sharing_2016}, HMDB \cite{wishart2018hmdb}, or large compound libraries such as PubChem \cite{kim_pubchem_2016}. Relying on databases inherently limits annotations to formulae that have already been observed. On the other hand, database independent (\textit{de novo}) chemical formula annotation considers all possible chemical formulae, though the task becomes more challenging due to the larger number of candidates. A recent bottom-up computational approach, BUDDY \cite{xing2023buddy}, is a hybrid of the two approaches and assigns database formulae to MS2 peaks and neutral losses. By combining peak and neutral loss annotations, BUDDY generates potential candidates not present in databases. However, overall annotation performance is a function both of how the candidate space is defined and an algorithm's ability to rank them; this limits our ability to perform a direct comparison to or analysis of BUDDY. Herein we focus specifically on \textit{de novo} formula annotation for maximal flexibility using only MS/MS information.

Notably, both MZMine \cite{ schmid2023integrative, pluskal2012highly} and  SIRIUS \cite{bocker2016fragmentation, bocker2009sirius, duhrkop_sirius_2019} have developed widely used methods for scoring chemical formula candidates using MS/MS information for \emph{de novo} annotation. While MZMine evaluates each candidate formula based on the number of MS/MS peaks it can explain,\cite{pluskal2012highly} SIRIUS scores them through a more expressive fragmentation tree strategy\cite{bocker2016fragmentation, bocker2009sirius, duhrkop_sirius_2019}. SIRIUS first proposes candidate formulae through an exhaustive enumeration step up to a certain mass error from the observed MS1, labels the MS2 peaks with potential ``subformulae'' of the candidate MS1 annotation, performs an optimization to arrange these subformula annotations into a fragmentation tree, and finally computes a likelihood of the chemical formula based upon the tree and isotopic patterns. Despite their success and widespread use, these methods offer room for improvement in terms of both accuracy and speed. We recently observed program timeouts using SIRIUS for larger molecules  (i.e., over 800 Da) during fragmentation tree calculations \cite{Goldman2022.12.30.522318}. An additional, lesser appreciated component of SIRIUS is that the method re-ranks candidate chemical formula based upon compound scores in the structure annotation step; this phase \emph{implicitly} reposes SIRIUS as a database dependent search and led to a nearly 50\% change in formula annotations in our previous study \cite{Goldman2022.12.30.522318}. It is unclear which pipeline steps lead to the high empirical formula annotation rates and the extent to which the tree score can be improved.  %

In this work, we present \methodNameFull{}, an energy-based modeling approach to improving the database-independent, \textit{de novo} chemical formula assignment step conditioned on both the MS1 mass and MS/MS spectrum. 
We previously demonstrated that Formula Transformers can be used to replace fragmentation tree \emph{kernels} in the annotation step.\cite{Goldman2022.12.30.522318} This study now demonstrates that fragmentation trees can be replaced throughout the MS/MS processing pipeline for equally accurate, fast, and robust predictions. As part of this, we develop a simple peak subformula assignment routine, thus circumventing the need for fragmentation tree construction. We rigorously evaluate the performance of \methodName{} on two datasets:  a public dataset subset from the GNPS \cite{wang_sharing_2016} we term \gnpsData{} \cite{duhrkop_systematic_2021, goldman2023prefix} and a variant of \gnpsData{} including spectra from the commercial \nistData{} dataset \cite{noauthor_tandem_nodate}. By training and evaluating our model both with and without data from \nistData{}, we enable reproducible evaluation of select trained models even in the absence of a \nistData{} license. \methodName{} achieves equivalent formula annotation accuracy on the positive mode CASMI2022 challenge spectra to the winning SIRIUS solution and outperforms the out-of-the-box SIRIUS assignments  by a margin of 18\% . Altogether, this demonstrates a path forward for replacing fragmentation trees in MS/MS data processing with a fully integrated deep learning processing structure annotation pipeline. 

We release \methodName{} as an open source tool that can be easily integrated into existing pipelines, with or without retraining, and is freely available under the MIT license at ref. Zenodo \zenodoIDGithub{} and \codeUrl{}.

\section{Methods}
\label{sec:methods}

\subsection{Preliminaries}
In an MS/MS experiment, an input small molecule $\mol$ is ionized, often by the addition of an adduct (denoted by one-hot vector, $\preAdduct$), measured, and fragmented in order to produce an MS/MS spectrum $\spec$. 
The spectrum is composed of peaks that can each be represented as mass, intensity pairs: 
\begin{equation}
    \spec \coloneqq \{ (\massIdx{1}, \intenIdx{1}), (\massIdx{2}, \intenIdx{2}), \dots (\massIdx{|\spec|}, \intenIdx{|\spec|}) \}
\end{equation}
Herein  we consider only positively charged ions and peaks and assume that each fragment carries only a single positive charge; m/z and mass are used interchangeably when describing observed peaks. 

In addition to the observed MS/MS spectrum $\spec{}$, we also have access to the MS1 measurements denoted as the precursor mass $\mass{}$. The core challenge of formula annotation is to determine the chemical formula of $\mol$ given $\mass$ and $\spec$. We denote the target chemical formula as a vector of integers, $\preForm{}$.

We represent the target chemical formula as a vector $\preForm{}$, in which each index $(\preForm)_{j} \in \intsPos$  corresponds to the integer count for the observed $j^{th}$ element, where the parenthetical denotes an index into the vector. In total, we consider a set of common elements, ``C'', ``N'', ``P'', ``O'', ``S'',
``Si'', ``I'', ``H'', ``Cl'', ``F'', ``Br'', ``B'', ``Se'', ``Fe'', ``Co'', ``As'', ``Na'', and ``K'' for formula vectors of size 18.  We consider common positive mode adducts when generating chemical formula candidates: \ce{[M\text{+}H]+}, \ce{[M\text{+}Na]+}, 
 \ce{[M\text{-}H2O\text{+}H]+}, \ce{[M\text{+}NH4]+}, \ce{[M]+}, and \ce{[M\text{-}{2}H2O\text{+}H]+}. The adduct candidate is specified by a one-hot vector $\preAdduct$.  For \emph{de novo} chemical formula annotation, $\mass$ can be decomposed into potential formula and adduct candidate options exhaustively using a highly efficient dynamic programming algorithm \cite{bocker2009sirius} within small 1 to 5 parts-per-million (ppm) mass tolerances. The generated (formula, adduct) candidates $\{(\preFormIdx{1}, \preAdductIdx{1}),(\preFormIdx{2}, \preAdductIdx{2}), (\preFormIdx{k}, \preAdductIdx{k})\}$ can be further filtered based on presence in a database or using various heuristics such as number of ring double bond equivalents if desired \cite{pretsch2000structure}.

In addition to inferring the chemical formula for the full molecule, a step common to SIRIUS fragmentation tree generation \cite{bocker2016fragmentation} and CSI:FingerID \cite{duhrkop_searching_2015} is to derive corresponding subformula annotations for the MS/MS peaks. In this way, a spectrum can be considered as a set of (subformulae, intensity) pairings, in which the set of peak masses $\{\massIdx{1}, \massIdx{2}, \dots \massIdx{|\spec|}\}$ is replaced with a set of chemical formula vectors $\{\prodFormIdx{1}, \prodFormIdx{2}, \dots, \prodFormIdx{\numAnnotated}\}$, where each chemical formula is a subset of the precursor formula, i.e., $\prodFormIdx{j} \subseteq \preForm{}$; peaks for which no formula can be assigned are excluded from the list. Because the fragment peaks are also charged, the observed masses are a summation of the mass of the subformula \emph{and} its adduct. We define the MS2 peak adduct for the $i^{th}$ peak  $\prodAdductIdx{i}$, the mass of which must be subtracted when considering chemical formula assignment. The parts-per-million difference at each peak between the adduct-adjusted peak mass and assigned subformula is referred to as $\prodFormError{i}$ to indicate measurement error.

\subsection{Approaching chemical formula annotation  with an energy based model formulation}
\label{sec:ebms}
As exhaustive chemical formula candidate generation can be solved via dynamic programming, \cite{bocker2009sirius} the key challenge in \emph{de novo} formula annotation is to \emph{score} how well each candidate formula matches the observed MS1 and MS2 spectra. Taking a probabilistic lens and following previous work in metabolomics and proteomics,\cite{bocker2009sirius, zhang2002probid} our goal is to learn the probability of a candidate chemical formula and adduct, $p_\theta(\preFormIdx{i}, \preAdductIdx{i} | \spec, M)$. We assume the MS1 is useful only for candidate generation, and so the problem is simplified to approximating $p_\theta(\preFormIdx{i}, \preAdductIdx{i} | \spec)$.

Energy based models (EBM) are a probabilistic modeling framework drawing inspiration from physics in which a probability distribution is defined by an  energy function, $E_\theta$.\cite{lecun2006tutorial} Mathematically, EBMs take the form:

\begin{equation}
p_{\theta}(x) = \frac{e^{-E_\theta(x)}}{Z(\theta)},
\end{equation}

\noindent where the denominator $Z(\theta) = \int_{x} e^{-E_\theta(x)}dx$ is referred to as a partition function and serves to normalize the energy to a valid probability, but is typically intractable to evaluate exactly. EBMs have reemerged in recent years, with work across the chemical sciences for reaction prediction, \cite{lin2022improving} retrosynthesis,\cite{sun2020energy} and scoring protein side-chain positions.\cite{du2020energy} These models are naturally suited for ranking applications. %
In our case, we factorize the probability of the candidate formula conditioned on the spectrum as an EBM of the form:

\begin{equation}
    p_\theta(\preFormIdx{i}, \preAdductIdx{i} | \spec)  = \frac{e^{\modelGeneral{}(\preFormIdx{i}, \preAdductIdx{i}, \spec)}}{Z(\theta)},
\end{equation}

\noindent where $\modelGeneral{}(\preForm{}, \preAdduct{}, \spec{})$ defines an arbitrary neural network energy function that takes as input the candidate formula, adduct, and full fragmentation spectrum. %
For any differentiable $\modelGeneral$, the energy function can be learned via a softmax loss function, aggregated over minibatches, and minimized via stochastic mini batch gradient descent: 

\begin{equation}
    \mathcal{L}(\preFormIdx{\text{True}}, \preAdductIdx{\text{True}}, \spec) = -\log \frac{e^{\modelGeneral{}(\preFormIdx{\text{True}},  \preAdductIdx{\text{True}}, \spec)}}{ 
 e^{\modelGeneral{}(\preFormIdx{\text{True}}, \preAdductIdx{\text{True}}, \spec)} +  \sum_{k=1}^K e^{\modelGeneral{}(\preFormIdx{k}, \preAdductIdx{k},  \spec)}},
\end{equation}

\noindent where $\{(\preFormIdx{k},\preAdductIdx{k})\}_{k=1}^K$ defines a set of ``decoy'' formulae for which $\modelGeneral$ will learn to assign a low score. The trained model can be applied independently to each candidate formula at inference time to yield a ranked list of assignments. 

Our approach is conceptually similar to SIRIUS \cite{bocker2009sirius, bocker2016fragmentation, duhrkop_sirius_2019}, but differs in that SIRIUS uses a heuristic maximum a posteriori (MAP) estimator to score fragmentation trees. This requires manually setting parameters such as the frequency of observing various fragments in the data. By using a flexible energy function trained via supervised learning, \methodName{} does not require manual parametrizations nor necessitate fragmentation tree generation. SIRIUS also allows for the incorporation of isotopic information from the MS1 to help identify chemical formula candidates. We focus solely on the MS2-related score rather than isotopic MS1 information, as this can be subsequently added and is often excluded from entries in spectral database such as the GNPS\cite{wang_sharing_2016}.

\subsection{The \methodName{} architecture}

\methodName{} parametrizes the energy function using a Chemical Formula Transformer\cite{Goldman2022.12.30.522318} we denote as $\modelMIST{}(\preForm, \preAdduct, \spec)$. 
For the input $(\preForm{}, \preAdduct{}, \spec)$, we first attempt to label each spectrum peak $(M^{(i)}, I^{(i)})$ with a plausible chemical subformula $\prodFormIdx{i}$ such that the spectrum can be represented by a set of subformulae (i.e., $\prodFormIdx{1}, \prodFormIdx{2}, \dots \prodFormIdx{3}$). Peaks are sorted by intensity and a maximum number, $N_p$ are retained, set to $20$ by default. \methodName{} then subsequently encodes these subformula peaks with a Chemical Formula Transformer, a Set Transformer variant \cite{ying_transformers_2021, Vaswani2017-se} we previously defined as part of the MIST architecture.\cite{Goldman2022.12.30.522318} The full \methodName{} pipeline is illustrated in Figure \ref{fig:overview}b, compared side-by-side to the SIRIUS model that requires a fragmentation tree calculation. We review the exact structure of this model below by describing each modeling step.

\paragraph{Subformula annotation} The first step in scoring a candidate formula for a spectrum is to assign a subformula to as many MS2 peaks as possible.
We begin by subtracting the precursor adduct mass from all mass values in the MS2. This approach assumes that the subpeaks have the same adduct ionization as the precursor ion, but could be modified in future iterations of \methodName{}. All  possible subformulae are enumerated and filtered with a ring double bond equivalent heuristic\cite{pretsch2000structure} to remove implausible formulae and generate a candidate set $\{\prodFormIdx{i}: \prodFormIdx{i} \subseteq \preForm, \text{RDBE}(\prodFormIdx{i}) \geq 0\}$.  The mass for each subformula is compared to all adduct-adjusted spectrum masses and the peak masses are assigned their nearest subformula match within a $15$ ppm tolerance.  An important contribution of this work with respect to the original MIST model \cite{Goldman2022.12.30.522318} is that this subformula assignment is now achieved with a compact and open source NumPy\cite{harris2020array} module, reducing the reliance on SIRIUS for high quality MS2 subformula annotations.

\paragraph{Formula embeddings} The subformula-annotated peaks are treated as integer vector inputs. Rather than pass these into a neural network directly, we first featurize each integer vector count into a sinusoidal embedding vector\cite{goldman2023prefix, Tancik2020-yy} and concatenate the resulting output. 
Briefly, 
 each integer element count, $v$, in the input formula is encoded by our counts-based encoder into the vector:
\begin{equation}
\text{ReLU} \left( \left[ \sin\left( \frac{2\pi v}{T_1} \right), \sin\left( \frac{2\pi v}{T_2} \right), \sin\left( \frac{2\pi v}{T_3} \right), \ldots \right] \right),
\end{equation}
where the periods ($T_1$, $T_2$, etc.) are set at increasing powers of two up to $256$ to discriminate all possible element counts given in the input, and $\text{ReLU}(\cdot)$ is the rectified linear unit activation function that results in only non-negative embeddings. All individual integer encodings within a formula are flattened and concatenated, leading to a full formula encoding  $\enc{\preForm} \in \realsPosCt{144}$.

\paragraph{Spectra context} In addition to the encodings, the  instrument type is considered as a one-hot vector covariate distinguishing between Ion Trap, Q-ToF, Orbitrap, and FTICR instruments. We denote this instrument type one-hot as $\specContext$.

\paragraph{Transformer architecture} Each peak subformula is first encoded into a hidden state vector that is then passed into the Transformer, with the precursor formula included as a special formula annotation $\prodFormIdx{0} \coloneqq \preForm{}$. In addition to encoding the formula, we add additional features for the formula \emph{difference} between the MS1 formula and each MS2 formula candidate vector (also converted into a sinusoidal embedding), a floating point scalar value for the MS2 peak intensity (set to 1 for the MS1 precursor formula candidate), the observed mass error between the observed adduct-adjusted mass and the monoisotopic mass of the MS2 subformula candidate (set to 0 for the MS1 precursor formula candidate), a one-hot encoding for the adduct type (assumed to be the same within a single spectrum)%
, the instrument type one-hot vector, a Boolean flag set to $1$ only at the precursor MS1 formula, and the number of total annotated peaks with subformula. These representations are embedded into hidden dimension $d$ with a shallow single hidden layer MLP:

\begin{equation}
\hiddenStatePrime{i} = [\enc{\prodFormIdx{i}}, \enc{\preForm{} - \prodFormIdx{i}}, \intenIdx{i}, \prodFormError{i}\indicator{\prodFormIdx{i} \neq \preForm}, \prodAdductIdx{i}, \specContext{}, \indicator{\prodFormIdx{i} = \preForm}, \numAnnotated{}]
\end{equation}
\begin{equation}
    \hiddenState{i} = \MLP{\hiddenStatePrime{i}}
\end{equation}

A standard multi-head Transformer, with a slight modification to include featurized attention between peaks as described previously in the MIST architecture \cite{Goldman2022.12.30.522318}, is then used to transform these peak representations into a score: 

\begin{equation}
\modelMIST(\preForm, \preAdduct{}, \spec) = \MLP{}\left(\transformer\left( 
   [\hiddenState{0}, \hiddenState{1},
    \dots \hiddenState{\numAnnotated{}}]
   \right)   \right)
\end{equation}

\noindent The representation from the Transformer is pooled into a fixed length vector by selecting the output representation at only the special $\hiddenState{0}$ position. Due to certain training set examples not including MS1 masses, we take care to avoid inputting the relative mass difference between the assigned precursor formula and MS1 mass which is artificially set to $0$. This helps avoid additional machine bias, as many of the training spectra have theoretical MS1 mass, rather than a measured MS1.

\paragraph{Model training} All models are trained using the aforementioned loss calculated using decoy formula. We sample minibatches of $b$ spectra, where there are up to $k=32$ decoys sampled for each spectrum in the minibatch, resulting in a total of $O(bk)$ candidate formula encoded per batch. To sample the most plausible ``hard'' negative decoys in each batch, we utilize a \fastFilter{} module described in detail below. All models are implemented in Python version 3.8 using PyTorch version 1.9 and PyTorch Lightning\cite{Falcon_PyTorch_Lightning_2019} version 1.6 and trained with the Adam optimizer.\cite{kingma2014adam}. Hyperparameters are optimized with Ray Tune\cite{liaw2018tune} version 2.0. Each model was trained on a single NVIDIA RTX A5000 GPU with training times taking under 3 hours of wall time. 

\noindent

\subsection{Generating formula candidates}
By default, \methodName{} utilizes the mass decomposition algorithm embedded within the SIRIUS software\cite{duhrkop_sirius_2019, bocker2007fast, duhrkop2013faster} (independent module from the rest of their pipeline) to decompose MS1 precursor masses into candidate formula options. We do not restrict the number of common elements C, N, O, or H. We set the maximum number of S and P atoms to 5 and 3, respectively, and limit each halogen (i.e., F, Cl, Br, I) to a maximum of one per formula, allowing for the recovery of 96\% of formulae in the test set. This constraint can be changed at inference time as desired. %
The chemical filter option ``COMMON'' is used to generate formulae for energy-based model training, and the ``RDBE'' filter is applied during inference. The default mass deviation is set to 10 ppm during training for all spectra. At inference, we vary the ppm tolerance for different instrument types (Ion Trap: 15 ppm; Q-ToF: 10 ppm; Orbitrap/FTICR: 5 ppm) as in BUDDY \cite{xing2023buddy}. We utilize SIRIUS version 5.6.3.

During inference, the user can choose to avoid this step in if they have their own, more narrow list of potential formulae candidates (e.g., generated by database search, knowledge about the chemical space being measured, or external tools such as BUDDY). In such cases, the exhaustive formula candidate generation step can be skipped and \methodName{}'s energy function can be used to directly evaluate the input candidate list.

\subsection{Downselecting candidate formulae with a learned filter}
Due to its energy-based formulation, the computational cost of \methodName{} scales linearly with the number of candidate precursor formulae. The evaluation of each formula requires assigning subformulae to peaks within a fragmentation spectrum, which is far faster than inducing a tree structure over subformulae but not negligible. %
To limit the space of candidate formulae, in addition to utilizingheuristics such as COMMON or RDBE as mentioned earlier, %
we train a data-driven filter we term \fastFilter{} (Figure \ref{fig:filter}a) to further narrow the option set. 

$\modelFast{}$ is a feedforward neural network that takes as input an encoded precursor formula candidate, $\enc{\preForm}$, and learns, in the same fashion as \methodName{}, to predict an energy value based solely on the formula---\emph{not} information about the spectrum---to approximate a non-normalized likelihood $p_\theta(\preFormIdx{i}|M)$. Because no spectrum information is needed, we train a single \fastFilter{} model using a large database of molecules and chemical formulae extracted from varous sources such as PubChem\cite{kim_pubchem_2016} and prepared by Duhrkop et al. \cite{duhrkop_systematic_2021} To avoid dataset bleed, all chemical formulae that appear in spectral libraries used for model training/testing are excluded. We use $\modelFast{}$ to select the top k candidates during inference (256 by default), or decoys during training, to further score with $\modelMIST{}(\preForm, \preAdduct, \spec)$.

\subsection{Datasets}

We evaluate \methodName{} in terms of its ability to predict precursor chemical formulae for MS/MS spectra from \gnpsData{}, a public natural products dataset extracted from the GNPS database\cite{wang_sharing_2016}. 
\gnpsData{} is prepared as in \citeauthor{goldman2023prefix}\cite{goldman2023prefix} and extracted from Duhrkop et al. \cite{duhrkop_systematic_2021} The dataset contains positive mode MS/MS spectra for compounds under 1,500 Da containing a predefined set of elements and adducts. In total \gnpsData{} contains 10,709 spectra, 8,553 unique structures and 5,433 unique chemical formulae. We employ chemical formula splits in which 10\% of the 5,433 chemical formulae are selected randomly and added---with their corresponding spectra---to the test set. 10\% of the remaining data is used for validation and early stopping.

In addition to the public data, we use the commercial \nistData{} library\cite{noauthor_tandem_nodate} to supplement the training dataset. We extract all Orbitrap high resolution positive mode mass spectra containing common elements and adducts. Examples with chemical formulae found in the test set are excluded to avoid biasing the model. In total, the combined dataset has 45,838 unique spectra, 30,950 unique 2D molecular structures, and 15,315 unique chemical formula. By using identical public test sets, we are able to report performance metrics with and without commercial library inclusion to enable replication studies and future methodological improvements.

\subsection{Baseline models}

In addition to learning $\modelGeneral$ using the MIST architecture, we select three separate baseline neural architectures: a feed forward network (FFN) inspired by MetFID \cite{fan_metfid_2020} that acts on a binned representation of the spectrum, $\modelFFN{}$; `MS1 Only'', a variant of $\modelFFN{}$ in which the binned spectrum is set to $\bm{0}$ for all spectra; and a Transformer model that utilizes multiscale sinusoidal embeddings  (Transformer) \cite{voronov_multi-scale_2022}, $\modelXformer{}$.  These models are trained and hyperparameter optimized equivalently to \methodName{}. In the FFN baseline, the binned spectrum representation is concatenated to the encoded MS1 candidate formula, the one-hot adduct representation, and the context vector, then passed into a multilayer perceptron module. The Transformer baseline concatenates the MS1 candidate formula encoding and context vector to the sinusoidal embedding at each of the top $100$ most intense peaks, along with the intensity. An additional ``cls'' token is added to the peaks containing the mass of the MS1 candidate and intensity of $2$. These are subsequently passed into a set of multi-head attention Transformer layers. The output is pooled at a special ``cls'' token. A single linear layer then predicts a scalar energy value.

\section{Results}

\subsection{\fastFilter{} limits the number of candidate formulae for \methodName{} consideration with high precision}

\begin{figure}[H]%
\centering
\includegraphics[width=1\textwidth]{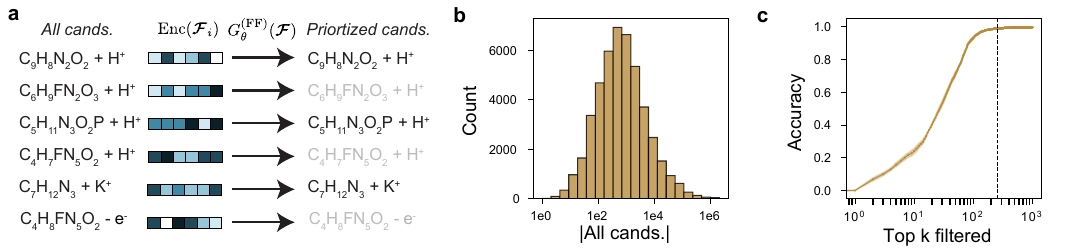}
\caption{\textbf{Large numbers of formula candidates can be quickly filtered with a simple feed forward neural network, \fastFilter{}.} \textbf{a.} Generated candidate formulae for a spectrum can be prioritized with a learned model. A scalar score is generated for each candidate formula given an MS1 value and mass tolerance via a feed forward neural network. %
Only the top ranked formulae are selected for consideration with \methodName{}.  \textbf{b.} The distribution of the number of candidates for all recorded MS1 spectra in our spectra libraries \gnpsData{} and \nistData{}. All candidate chemical formulae are generated for all 6 adduct types considered with an allowed mass deviation of 10 ppm. \textbf{c.} \fastFilter{} model accuracy at various top k cutoff values. The top formula is nearly always recovered within the top 256 candidates. The 95\% confidence interval of the mean for recovery is shown across the 3 different formula splits considered in this work. All results are computed for a single trained \fastFilter{} model on a large database of biologically-relevant molecules.
}\label{fig:filter}
\end{figure}

A key limiting factor in \textit{de novo} formula identification is the large size of the candidate space, particularly for molecules with higher masses. %
We first quantify the size of the candidate space. We process the full set of training spectra, including both \gnpsData{} and \nistData{}, into chemical formula candidates with the permissive RDBE filter. Over 15\% of spectra have greater than 5,000 formula candidates (Figure \ref{fig:filter}b). 
By training a light-weight model, \fastFilter{},  to predict how likely each formula is to appear in a biologically-relevant database of small molecules,  we are able to filter the candidates down to a smaller subset. 
We are able to recover the true formula 99\% of the time within the top 256 candidates (Figure \ref{fig:filter}c). This guarantees the computational tractability of the \methodName{} pipeline, as subformula labeling would be prohibitively expensive for spectra with hundreds of thousands or millions of candidate formulae. %

\subsection{Chemical Formula Transformers provide meaningful representations for formula annotation }
\begin{figure}[H]%
\centering
\includegraphics[width=1\textwidth]{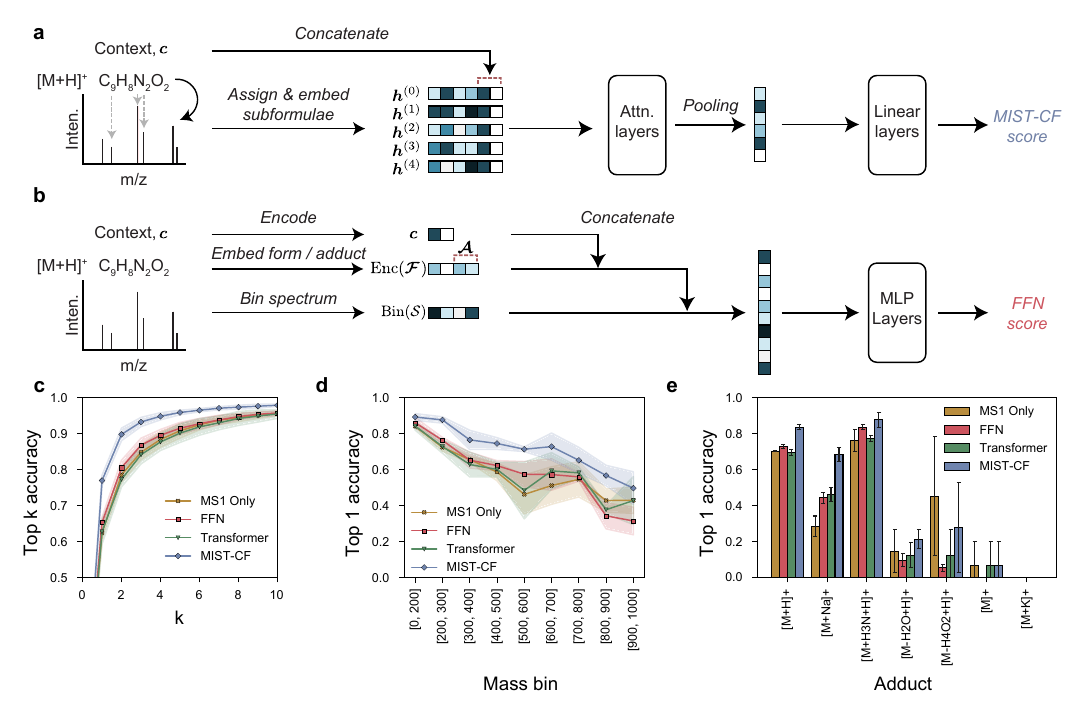}
\caption{\textbf{\methodName{} is a highly-effective architecture for learning to rank plausible MS1 formula annotations} \textbf{a.} The \methodName{} architecture uses the candidate chemical formulae to generate subformulae and encode these with a Formula Transformer. \textbf{b.} The baseline feed forward network (FFN) separately encodes the spectra and formula before feeding their concenation into a multilayer perceptron (MLP). \textbf{c.}  For all spectra in the test set, the fraction recovered at various top k values for all methods is computed and shown. \textbf{d-e.} The top 1 accuracy for the methods is grouped by the mass of the MS1 precursor or adduct type. All results are computed for \methodName{} and the following baselines: MS1 only (a feed forward network utilizing only the chemical formula and context vector), a feed forward network, and a Transformer. All results are computed over 3 random formula splits and respective training runs as described in \nameref{sec:methods}, where the \nistData{} is included in the training sets. All spectra include up to 256 candidates to select from as selected with the ``COMMON'' filter \cite{duhrkop_sirius_2019} and \fastFilter{}. Error bars and shaded regions show 95\% confidence intervals of the mean.
}\label{fig:benchmark}
\end{figure}

We evaluated four different architectures for encoding spectra and formulae to determine the best neural network architecture for the energy-based modeling framework described in \nameref{sec:methods}. \methodName{} uses a Transformer model to convolve upon MS2 subformula with context information concatenated to each formula (e.g., adduct, instrument type) (Figure \ref{fig:benchmark}a). We compare this model to a feed forward neural network (Figure \ref{fig:benchmark}b), a standard Transformer applied to sinusoidal embeddings of each m/z value \cite{voronov_multi-scale_2022}, and a variant of the feed forward network that does not embed the spectrum (``MS1 only''). Importantly, all baseline neural network models are provided with the same context vectors, training set, and hyperparameter optimization schemes to enable fair comparison. 

We find that \methodName{} outperforms all other architectures tested, likely due to its ability to explicitly combine the representation of the spectrum and formula candidate, rather than via intra-architecture concatenations (Figure \ref{fig:benchmark}b). We test the model by holding out a number of spectra and chemical formulae from model training. For each test spectrum, up to $256$ formula and adduct candidates are re-ranked by each model. \methodName{} outperforms the next best model at top 1 accuracy by a 10\% margin (Figure \ref{fig:benchmark}c; Table \ref{tab:benchmark}). 
Curiously, the feed forward network and Transformer network show little difference from each other, indicating that neural network architecture (i.e., FFN vs. Transformer) has less impact than the input information provided to the model (i.e., subformula labels)---the main strength of \methodName{}. All models are able to perform at least as well as the MS1 only model, consistent with our intuition that the models are learning more than just database bias. 

All models decrease in accuracy as the masses of compounds increase (Figure \ref{fig:benchmark}d) due to the growing number of plausible candidates. Similarly, models struggle on  adducts that appear less often in the training dataset, such as potassium adducts (Figure \ref{fig:benchmark}e). This is one such area where manually parametrized models such as SIRIUS \cite{duhrkop_sirius_2019} may be better suited to generalization. Data-driven methods such as \methodName{} are empirically better at correctly retrieving formulae with common proton or sodium adducts.

Integrating the higher quality Orbitrap training spectra from \nistData{} leads to improved performance on the same test set. \methodName{} models trained on \gnpsData{} alone achieve a top 1 accuracy of 0.741 compared to a top 1 accuracy of 0.769 for models that train on the \nistData{} in addition (Table \ref{tab:benchmark}). This $>2\%$ absolute improvement underscores that formula prediction accuracy may be enhanced with the upcoming release of  new spectral libraries and higher quality data.\cite{bittremieux2022critical}

\begin{table}[ht]
\centering
\caption{\methodName{} outperforms comparable neural network baselines at chemical formula annotation from MS/MS. Models were trained to predict held out \gnpsData{} test examples using training sets consisting of ``\gnpsData{}'' or ``\gnpsData{} + \nistData{}.'' The best value in each column is typeset in bold.  Models were evaluated using three independent formula splits. Values are shown $\pm$ standard errors of the mean.}
\resizebox{\textwidth}{!}{
\begin{tabular}{lllllll}
\toprule
Training dataset & \multicolumn{3}{c}{\gnpsData{}} & \multicolumn{3}{c}{\gnpsData{} + \nistData{}} \\
\cmidrule(r){2-4} \cmidrule(r){5-7}
Top k &                 \hfil 1 &                \hfil  2 &               \hfil   3 &           \hfil       1 &             \hfil     2 &      \hfil            3 \\
\midrule\midrule
MS1 Only    &  $0.609 \pm 0.002$ &  $0.773 \pm 0.003$ &  $0.833 \pm 0.003$ &  $0.623 \pm 0.007$ &  $0.785 \pm 0.008$ &  $0.847 \pm 0.008$ \\
FFN         &  $0.635 \pm 0.009$ &  $0.786 \pm 0.009$ &  $0.847 \pm 0.007$ &  $0.652 \pm 0.006$ &  $0.804 \pm 0.009$ &  $0.867 \pm 0.008$ \\
Transformer &  $0.639 \pm 0.006$ &  $0.791 \pm 0.009$ &  $0.851 \pm 0.006$ &  $0.626 \pm 0.014$ &  $0.772 \pm 0.008$ &  $0.840 \pm 0.011$ \\
\midrule\midrule
\methodName{}&  \bm{$0.741 \pm 0.010$} &  \bm{$0.878 \pm 0.009$} &  \bm{$0.919 \pm 0.007$} &  \bm{$0.769 \pm 0.006$} &  \bm{$0.897 \pm 0.007$} &  \bm{$0.931 \pm 0.005$} \\
\bottomrule
\label{tab:benchmark}
\end{tabular}
}
\end{table}

\begin{figure}[H]%
\centering
\includegraphics[width=0.5\textwidth]{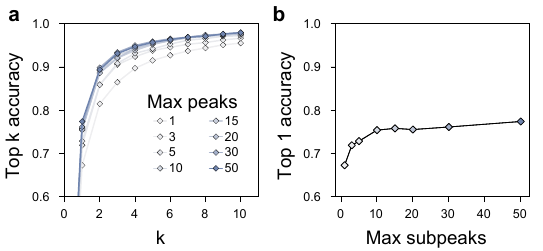}
\caption{\textbf{MS/MS subpeaks drive \methodName{} performance.} \textbf{a.} The (sub)Formula Transformer in \methodName{} %
operates on only the $\numAnnotated{}$ highest intensity MS2 peaks. Here, from that set of labeled MS2 peaks, the number of peaks used as input to the spectrum is limited. As the number of included peaks increases, the formula retrieval accuracy does too. %
\textbf{b.} The top 1 retrieval accuracy is shown for all maximum subpeak numbers. All results are shown for \methodName{} trained on the joint \gnpsData{} and \nistData{} dataset and tested on a single test split of the data. 
} 
\label{fig:subpeaks}
\end{figure}

One of the core methodological decisions in \methodName{} is the application of a Formula Transformer to the set of labeled subformulae in the MS2 spectrum. To verify that these peaks are informative for the model, we repeated benchmarking experiments for a single split of the data, this time modulating the maximum number of peaks (rank-ordered by intensity) viewable by \methodName{}. 

Model performance rapidly increases as the maximum number of included peaks increases, sharply rising from $0.662$ to $0.720$ for $1$ and $2$ spectrum peaks respectively, where $\numAnnotated{}=0$ is functionally equivalent to the MS1 Only baseline with an accuracy of $0.623$ (Table \ref{tab:benchmark}). There are diminishing returns of including lower intensity peaks, with $1$, $5$, $10$,  $20$, $50$ maximum formula peaks achieving top 1 accuracies of  $0.673$, $0.729$, $0.754$, $0.756$, and $0.774$ respectively (Figure \ref{fig:subpeaks}). Making the model aware of a greater number of fragments enables more generalizable and accurate predictions. This builds confidence that the model is learning more than database biases, drawing information from even minor peaks to inform predictions. As can be seen, the absolute difference in performance appears to level off beyond $20$ with only more marginal increases toward $\numAnnotated{} = 50$. Given that the performance benefit is marginal for many more peak annotations, we maintain our peak default $\numAnnotated{} = 20$ unless otherwise stated to reduce the runtime of the method.

\subsection{\methodName{} compares favorably to existing formula annotation tools}

\begin{figure}[H]%
\centering
\includegraphics[width=1\textwidth]{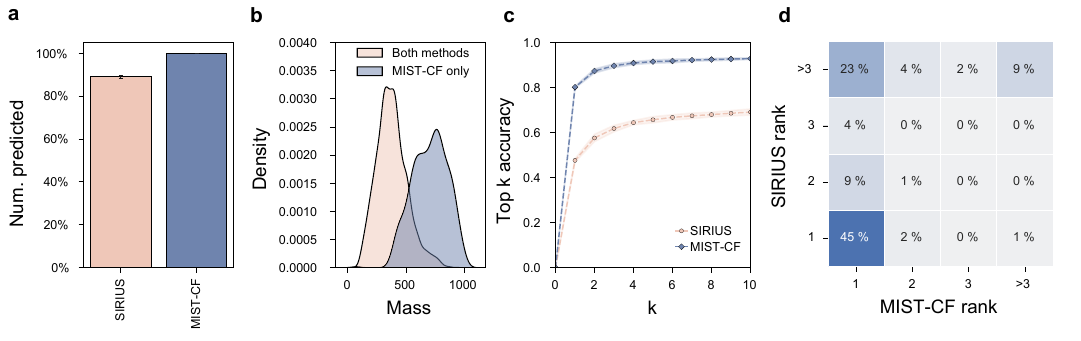}
\caption{\textbf{\methodName{} more frequently assigns correct chemical formulae than the SIRIUS formula assignment module on the \gnpsData{} test set.}  %
\textbf{a.} The number of spectra for which each method is able to predict a formula---regardless of accuracy. \textbf{b.} The distribution of molecular masses for the spectra both methods are able to annotate---regardless of accuracy--- compared to the distribution of molecular masses for spectra on which only \methodName{} succeeds. \textbf{c.} Top k accuracy for both methods is shown. \textbf{d.} The rank at which each method is able to recover the true chemical formula. This visualization highlights that there are many spectra for which \methodName{} achieves rank 1 that SIRIUS does not predict in its top 3. \methodName{} models are trained on the joint \nistData{} and \gnpsData{} dataset. %
SIRIUS is executed with a compound and tree timeout of 300 seconds. %
All values are shown for 3 random formula splits of the \gnpsData{} data. Error bars and confidence intervals show 95\% confidence intervals for the standard error of the mean. Accuracy is computed with respect to the \emph{total} MS1 composition (i.e., summed adduct and formula) to avoid biasing results against SIRIUS.
}\label{fig:sirius_retrospect}
\end{figure}

The widely used SIRIUS tool \cite{duhrkop_sirius_2019} is the \emph{de facto} state of the art for the task of chemical formula annotation. %
We therefore perform  a head-to-head comparison of \methodName{} and SIRIUS on the same \gnpsData{} test set. Herein, \methodName{} considers all candidate MS1 formulae within 10 ppm of the recorded MS1 mass and utilizes \fastFilter{} to down select to 256 candidates for full evaluation.  
We run the SIRIUS formula module from the command line with tree and compound timeouts of 300 seconds to avoid excessive execution times.  With these constraints, \methodName{} is able to  predict a formula for every spectrum, whereas SIRIUS fails for 10.77\% of the spectra (Figure \ref{fig:sirius_retrospect}a). Failed spectra are associated with larger masses on average (700.80 Da) than successful ones (378.06 Da).

Model accuracy is evaluated in terms of the summed elemental composition of the formula and adduct. SIRIUS does not distinguish tree scores with different adduct assignments as we do in \methodName{} by design (e.g., [\ch{C6H12O6}+\ch{H}]\textsuperscript{+} and [\ch{C6H10O5}+\ch{H}-\ch{H2O}]\textsuperscript{+} would appear to have equivalent tree scores).  On the \gnpsData{} test set, \methodName{} successfully predicts chemical formulae with a 71\% top 1 accuracy, compared to SIRIUS's 48\% accuracy. This represents a  >20\% improvement in absolute top 1 prediction accuracy (Figure \ref{fig:sirius_retrospect}c). On a per-spectrum basis, SIRIUS rarely makes correct predictions \methodName{} does not, with a total of approximately 3\% of spectra falling into this category, whereas \methodName{} appears to predict 36\% of spectra correctly when SIRIUS cannot (Figure \ref{fig:sirius_retrospect}d).

Beyond the improvement in accuracy, \methodName{} required approximately one-third the wall time of SIRIUS (evaluated on compounds under 700 Da for which SIRIUS does not time out) when both methods are run on a single CPU core (Supplementary Information). %

\subsection{ \methodName{} achieves competitive out-of-the-box performance on the CASMI2022 challenge}

We selected the dataset released as part of the Comparative Assessment of Small Molecule Identifications 2022 (CASMI2022) \cite{noauthor_critical_nodate} for additional validation. %
Our training datasets were generated prior to the competition announcement, minimizing the risk of training dataset bias and simulating a prospective use case. %
Herein, we focus only on the formula identification challenge rather than the full task of structural elucidation.

We extract all positive mode MS/MS files from the provided mzML files using MZMine 3 \cite{schmid2023integrative} and apply both \methodName{} and SIRIUS to predict chemical formulae for each of the extracted 304 spectra. Using our constraints on element types, 296 of the 304 formulae are recoverable (97\%), excluding species such as iodixanol (\ch{C35H44I6N6O15}). We use an MS1 tolerance of 5 ppm.

Model performance was evaluated on four key metrics: the accuracy of predicting formula correctly, the accuracy of predicting the adduct correctly, and the accuracy of predicting the total elemental composition of the formula and adduct combined. As discussed above, we use this last metric because evaluating the accuracy of the chemical formula alone would bias the comparison for \methodName{}.

We consider three variants of SIRIUS: SIRIUS, SIRIUS (CSI:FingerID) and SIRIUS (Submission). SIRIUS (CSI:FingerID) provides formulae as re-ranked by their CSI:FingerID score when searched against PubChem. SIRIUS (Submission) uses predictions submitted by the SIRIUS authors at the latest competition under file name ``duehrkop\_CASMI2022.csv.'' This submission reportedly used a mix of ion identity molecular networking \cite{schmid2021ion}, which can be used to more accurately resolve adduct types, along with manual curation to improve performance.

Encouragingly, we find that \methodName{} is competitive with SIRIUS (Submission), achieving an equivalent collective joint formula and adduct accuracy of 0.868, despite our automated classification and no additional manual curation (Table \ref{tab:casmi22_acc}). If evaluating SIRIUS (Submission) on the subset of 272 spectra for which they submitted a prediction, a higher accuracy is achieved compared to  \methodName{}, as \methodName{} predicts \emph{all} test set spectra. Nevertheless, when using default parameters, \methodName{} reaches a top 1 formula accuracy of 0.822 (including adduct prediction) compared to an accuracy of 0.516 for SIRIUS (CSI:FingerID). These results illustrate the competitiveness of \methodName{} and demonstrate that accurate, prospective formula annotation does not require computing full fragmentation trees.  In addition to testing our default \methodName{} model, we also test a variant with $\numAnnotated{}=50$ as a comparison. Performance is equivalent for the joint accuracy of formula and adduct pairs. However, accuracy in predicting the formula alone increases, as the model appears to be marginally better at adduct assignment.

\begin{table}[ht]
\centering
\caption{Model accuracy of \methodName{} and SIRIUS evaluated on CASMI2022. %
``Accuracy ($\preForm{}$)'' indicates accuracy of predicting the exact formula correctly. ``Accuracy ($\preAdduct{}$)'' indicates the fraction of spectra for which the first predicted (formula, adduct) pair includes the correct adduct. ``Accuracy ($\preForm{} + \preAdduct{}$)'' indicates the fraction of spectra for which the total elemental composition of the top predicted formula and adduct matches the elemental composition of the summed true formula and adduct. ``Predicted'' indicates the number of spectra for which a formula and adduct prediction was made. ``SIRIUS'' makes predictions with the default command line tool described above; ``SIRIUS (CSI:FingerID)'' predicts formula, adduct pairs using CSI:FingerID rankings against PubChem; SIRIUS (Submission) is taken directly from the previous CASMI2022 results. \methodName{} (20 peaks) uses default 20 subformula peaks, whereas \methodName{} (50 peaks) utilizes 50 subformula peaks. Accuracy ($\preForm{}$) and Accuracy ($\preAdduct{}$) are not reported for SIRIUS as these numbers are unfairly low due to how we break ties between predictions with equivalent scores. All accuracies are reported for the top 1 prediction and divided by the total number of spectra (304), not the total number of predicted spectra. Numbers typeset in bold indicate the best result in the column.}
\label{tab:casmi22_acc}
\resizebox{0.7 \textwidth}{!}{
\begin{tabular}{lrrrr}
\toprule
{Method} &  Accuracy &  Accuracy  &  Accuracy &   \\
{} &  ($\preForm$) &  ($\preAdduct$) &  ($\preForm+\preAdduct$) &  Predicted \\
\midrule \midrule
SIRIUS version 5.6.3                &               \-- &              \-- &                        0.641 &             274 \\
SIRIUS version 5.6.3 (CSI:FingerID)  &               0.516 &              0.543 &                        0.678 &             254 \\
SIRIUS (Submission)      &               \textbf{0.865} &              0.855 &                        \textbf{0.868} &             272 \\
\midrule\midrule
\methodName{} (50 peaks)               &               0.842 &              \textbf{0.901} &                        0.862 &             \textbf{304} \\
\methodName{} (20 peaks) &               0.822 &              0.885 &                        \textbf{0.868} &             \textbf{304} \\

\bottomrule
\end{tabular}
}
\end{table}

\section{Conclusion}

We have introduced a data-driven neural network model, \methodName{}, for inferring chemical formulae from MS/MS spectra, trained using an energy-based modeling framework. We benchmark this model extensively to show how our recent Chemical Formula Transformer architecture is uniquely suited to this task of integrating formula and spectra information. \methodName{} outperforms other learning-to-rank neural network architectures and ties the winning solution at the CASMI 2022 competition within the positive mode category, despite using zero MS1 isotopic information or manual prediction refinement.

This work defines a clear problem formulation for learning to rank MS1 candidate formula from MS/MS data, including open source code and data splits. To address this task, we release open source and trained models with low memory and run-time concerns, even for large molecules. In addition to these models, we develop a smaller and light-weight neural network formula filtering model to help prioritize biologically-relevant chemical formula candidates. This work continues to demonstrate the efficacy of our recent Chemical Formula Transformer network architecture in thorough benchmarking comparisons  \cite{Goldman2022.12.30.522318}. As part of this, we have greatly simplified and improved the the Chemical Formula Transformer implementation to now utilize custom formula embeddings, a simple and open source subformula assignment routine (i.e., no longer fragmentation trees), additional model inputs such as instrument type, and multiple adduct types. Applying these same changes to the Chemical Formula Transformer architecture for fingerprint prediction is likely to yield similar improvements.

There are many avenues for improving upon \methodName{}. We have trained models only for positive-mode data; we do not directly address the integration of \methodName{} scores with MS1 isotopic scores; we do not consider adduct switching in MS2 subformula assignment; we still rely upon SIRIUS's algorithmic decomposition of exact masses into formula candidates due to their fast implementation; and we have not yet explored the use of forward structure-to-spectrum models \cite{goldman2023generating, goldman2023prefix} for data augmentation.  There is also an opportunity to combine \methodName{} with the recently reported  BUDDY \cite{xing2023buddy} by re-ranking formulae generated by BUDDY rather than relying on the \fastFilter{}.

Altogether, we are optimistic about the potential to integrate this model into existing pipelines for small molecule metabolite identification. By addressing the task of chemical formula annotation, this work moves us one step closer to our vision of an integrated neural network driven metabolite annotation pipeline.

\section{Data and Software Availability}
All code to replicate experiments, train new models, and load pre-trained models is available at \codeUrl{}. The code to parse NIST can be found in \codeUrlNist{} but this optional subset of training data requires a license and cannot be shared publicly. The exact repository version used in this work has been archived at Zenodo record \zenodoID{} (data) and Zenodo record \zenodoIDGithub{} (code).

\section{Authors' contributions}
S.G. and J.X. jointly wrote the software and conducted experiments. J.P. assisted with CASMI2022 data processing and experiments. S.G., J.X., and C.W.C. conceptualized the project, designed model components, and wrote the manuscript. C.W.C supervised the work. 

\begin{acknowledgement}
The authors thank K. Duhrkop for helpful discussions around the SIRIUS method. S.G. was supported by the MIT-Takeda Fellowship program and the Machine Learning for Pharmaceutical Discovery and Synthesis consortium. J.P. was supported by the MIT-Accenture Fellowship. 

\end{acknowledgement}

\bibliography{refs}

\end{document}


\newpage

\begin{suppinfo}

\section{SIRIUS configuration}
\label{SI:sirius}
SIRIUS version 5.6.3  \cite{duhrkop_sirius_2019} is used as a baseline throughout the work. For consistency,  we utilize a 300 second (5 minute) timeout for each compound and set equivalent adducts and formula as used in MIST-CF. An example call signature using our parameters, as we use in CASMI2022 is shown in Listing \ref{lst:sirius_call}.

\small
\begin{lstlisting}[frame=single,language=bash,caption={An example call to SIRIUS for annotating  the formula for an input MGF file, \textsf{\$INPUT\_MGF}},label={lst:sirius_call}]
$SIRIUS_PATH  \
    --cores $CORES \
    --output  $OUTFOLDER1 \
    --input $INPUT_MGF \
    formula  \
    -i "[M+H]+,[M+K]+,[M+Na]+,[M+H-H2O]+,[M+H-H4O2]+,[M+NH4]+,[M]+" \
    -e "C[0-]N[0-]O[0-]H[0-]S[0-5]P[0-3]I[0-1]Cl[0-1]F[0-1]Br[0-1]" \
    --ppm-max 5.0 \
    --tree-timeout 300 \
    --compound-timeout 300 \
    fingerprint \
    structure \
    write-summaries \
    --output $OUTFOLDER2 
\end{lstlisting}
\normalsize

\section{Timing experiments}
\label{SI:Timing}
To demonstrate that \methodName{} has reasonable runtime properties, we conduct a simple timed experiment. 10 random sample test set spectra under $700$ Da from the \gnpsData{} were selected and set aside in an MGF file. Using the command line call signature for formula prediction on a Linux workstation, we test the wall time  to predict formula for these spectra with all pre-specified elements, a maximum ppm of 10 from the MS1, and all adduct possibilities considered within \methodName{}. We repeat the same procedure for SIRIUS, using a compound and tree timeout of 1 minute to avoid excessively large times. We conduct these experiments using only 1 CPU core. We find that in three separate trials, \methodName{} completed in $39.31 \pm 0.49$ seconds (mean $\pm$ standard error of the mean), whereas SIRIUS completed in $112.61 \pm 0.21$ seconds (mean $\pm$ standard error of the mean).

\section{Hyperparameter setting}
\label{SI:param}
To enable fair comparison across models, hyperparameters were tuned for
\methodName{}, the FFN binned prediction baseline, the transformer binned prediction baseline, and the \fastFilter{} model.  Parameters were tuned using RayTune \cite{liaw2018tune} with Optuna \cite{akiba2019optuna} and an
ASHAScheduler. Each model was allotted 50 different hyperoptimization trials for
fitting. \methodName{}, the FFN baseline, and the Transformer baseline were hyperparameter optimized on a smaller $10,000$ spectra subset
of the combined \gnpsData{} and \nistData{} datasets. \methodName{} was hyperparameter optimized using a maximum of $10$ peaks per spectrum. The \fastFilter{} model was hyperparameter optimized directly on the biological molecules dataset. Parameters are listed in Table \ref{tab:param}.

\begin{table}[H]
\caption{Hyperparameter settings for \methodName{}, FFN, Transformer, and the \fastFilter{} model to select batches of chemical formula.}
\label{tab:param}
\begin{center}
\renewcommand{\arraystretch}{1.0}
\begin{tabularx}{\textwidth}{llll}
\toprule
\textbf{Model}      &\textbf{Parameter}      &\textbf{Grid}   &\textbf{Value} \tabularnewline
\midrule
\methodName{}     &Learning rate  &[1e-4, 1e-3]     &0.00045       \tabularnewline
&Learning rate decay fraction   &[0.7, 1.0]   &0.8830        \tabularnewline
&Weight decay   &\{1e-6, 1e-7, 0.0\}  & 0.0   \tabularnewline
&Hidden size    &\{32, 64, 128, 256\}    &128      \tabularnewline
&Dropout     &[0, 0.5]  & 0.1   \tabularnewline
&Number of layers     &[1, 2, 3, 4] & 2     \tabularnewline
&Batch size     &\{4, 8, 16\} &4       \tabularnewline

\midrule
FFN &Number of bins  &\{1000, 2000, 3000, ..., 10000\}    &5000      \tabularnewline
&Learning rate  &[1e-4, 1e-3]     &0.0005      \tabularnewline
&Learning rate decay fraction   &[0.7, 1.0]   &0.8477  \tabularnewline
&Weight decay   &\{1e-6, 1e-7, 0.0\}  & 0    \tabularnewline
&Hidden size    &\{32, 64, 128, 256, 512\}   & 512    \tabularnewline
&Dropout     &[0, 0.5]  & 0.4   \tabularnewline
&Number of layers     &\{1, 2, 3, 4\} & 4     \tabularnewline
&Batch size     &\{16, 32, 64\} &64      \tabularnewline
\midrule
Transformer
&Learning rate  &[1e-4, 1e-3]     &0.00025     \tabularnewline
&Learning rate decay fraction   &[0.7, 1.0]   &0.8601 \tabularnewline
&Weight decay   &\{1e-6, 1e-7, 0.0\}  &1e-7  \tabularnewline
&Hidden size    &\{32, 64, 128, 256, 512\}   &32    \tabularnewline
&Dropout     &[0, 0.5]  & 0.2   \tabularnewline
&Number of layers     &\{1, 2, 3, 4\} &2 \tabularnewline
&Batch size     &\{8, 16\} & 8

\tabularnewline
\midrule
\fastFilter{}     &Learning rate  &[1e-4, 1e-3]     &0.0003     \tabularnewline
&Learning rate decay fraction   &[0.7, 1.0]   &0.8642      \tabularnewline
&Weight decay   &\{1e-6, 1e-7, 0.0\}  & 0   \tabularnewline
&Hidden size    &\{32, 64, 128, 256, 512\}   &256    \tabularnewline
&Dropout     &[0, 0.5]  & 0.1   \tabularnewline
&Number of layers     &\{1, 2, 3, 4\} &3     \tabularnewline
&Batch size     &\{16, 32, 64\} &64      \tabularnewline
\bottomrule
\end{tabularx}
\end{center}
\end{table}

\end{suppinfo}

\bibliography{refs}